\documentclass[11pt,preprint2]{aastex}

\newcommand{\msun}{{M$_\odot$}}
\newcommand{\mbh}{M_{\rm BH}}

\newcommand{\msig}{$M_{\rm BH}-\sigma_{\rm \ast}$}
\newcommand{\mlr}{$\Upsilon_{\rm R}$}

\def\min{\hbox{${}^{\prime}$}}
\def\sec{\hbox{${}^{\prime\prime}$}}

\begin{document}
\title{The Black Hole Mass
  of NGC~4151: Comparison of Reverberation Mapping and Stellar Dynamical
  Measurements}

\author{Christopher A. Onken\altaffilmark{1,2}, Monica Valluri\altaffilmark{3},
Bradley M. Peterson\altaffilmark{2}, Richard W. Pogge\altaffilmark{2},
Misty C. Bentz\altaffilmark{2}, Laura Ferrarese\altaffilmark{1}, 
Marianne Vestergaard\altaffilmark{4}, D. Michael Crenshaw\altaffilmark{5},
Sergey G. Sergeev\altaffilmark{6,7}, Ian M. McHardy\altaffilmark{8},
David Merritt\altaffilmark{9}, Gary A. Bower\altaffilmark{10},
Timothy M. Heckman\altaffilmark{11}, and Amri Wandel\altaffilmark{12}}
\altaffiltext{1}{Herzberg Institute of Astrophysics, 5071 West Saanich Road, Victoria, BC V9E 2E7, Canada}
\altaffiltext{2}{Department of Astronomy, The Ohio State University, 140 West 18th Avenue, Columbus, OH 43210}
\altaffiltext{3}{Kavli Institute for Cosmological Physics, University of Chicago, 5640 South Ellis Avenue, Chicago, IL 60637}
\altaffiltext{4}{Steward Observatory, University of Arizona, 933 North Cherry Avenue, Tucson, AZ 85721}
\altaffiltext{5}{Department of Physics and Astronomy, Astronomy Offices, Georgia State University, One Park Place South SE, Suite 700, Atlanta, GA 30303}
\altaffiltext{6}{Crimean Astrophysical Observatory, P/O Nauchny, Crimea 98409, Ukraine}
\altaffiltext{7}{Isaak Newton Institute of Chile, Crimean Branch, Ukraine}
\altaffiltext{8}{School of Physics and Astronomy, University of Southampton, Highfield, Southampton SO17 1BJ, UK}
\altaffiltext{9}{Department of Physics, Rochester Institute of Technology, 85 Lomb Memorial Drive, Rochester, NY 14623}
\altaffiltext{10}{Computer Sciences Corporation, Space Telescope Science Institute, 3700 San Martin Drive, Baltimore, MD 21218}
\altaffiltext{11}{Department of Physics and Astronomy, Johns Hopkins University, 3400 North Charles Street, Baltimore, MD 21218}
\altaffiltext{12}{Racah Institute of Physics, The Hebrew University, Jerusalem 91904, Israel}
\email{christopher.onken@nrc.gc.ca}

\begin{abstract}
  We present a stellar dynamical estimate of the black hole (BH) mass in the
  Seyfert~1 galaxy, NGC~4151. We analyze ground-based spectroscopy as well
  as imaging data from the ground and space, and we construct 3-integral
  axisymmetric models in order to constrain the BH mass and mass-to-light
  ratio. The dynamical models depend on the assumed inclination of the
  kinematic symmetry axis of the stellar bulge. In the case where the bulge
  is assumed to be viewed edge-on, the kinematical data give only an upper
  limit to the mass of the BH of $\sim 4\times 10^{7}$~M$_{\odot}$
  (1$\sigma$). If the bulge kinematic axis is assumed to have the same
  inclination as the symmetry axis of the large-scale galaxy disk (i.e.,
  23$^{\circ}$ relative to the line of sight), a best-fit dynamical mass
  between $4-5 \times 10^{7}$~M$_{\odot}$ is obtained. However, because of
  the poor quality of the fit when the bulge is assumed to be inclined (as
  determined by the noisiness of the $\chi^2$ surface and its minimum
  value), and because we lack spectroscopic data that clearly resolves the
  BH sphere of influence, we consider our measurements to be tentative
  estimates of the dynamical BH mass.  With this preliminary result,
  NGC~4151 is now among the small sample of galaxies in which the BH mass
  has been constrained from two independent techniques, and the mass values
  we find for both bulge inclinations are in reasonable agreement with the
  recent estimate from reverberation mapping ($4.57^{+0.57}_{-0.47}\times
  10^{7}$~M$_{\odot}$) published by Bentz et al.
\end{abstract}


\section{Introduction}

Evidence for the existence of supermassive black holes (BHs) in galactic
centers has progressively become stronger over the past decade
\cite[see][for a recent review]{ferrarese05}. Direct kinematical
measurements of BH masses ($\mbh$) from tracers of various types have been
made in roughly three dozen galaxies. This growth in the number of BHs with
mass measurements has led to the discovery of correlations between $\mbh$
and various properties of the host galaxies.  The measured masses have been
found to be correlated with bulge luminosity \citep{KR95, McLure02,
  marconi03, Erwin04}, bulge mass \citep{merritt01,marconi03,haering04}, the
central velocity dispersion of the stellar component \cite[the \msig\
relationship;][]{ferrarese00,gebhardt00a}, the degree of concentration of
bulge light \citep{Graham01, Erwin04, graham07}, and, possibly, the mass of 
the surrounding dark matter halo \citep{Ferrarese02}.

Most of the BHs that were used to define the \msig\ and other scaling
relationships between BH masses and host galaxy properties reside in
quiescent galaxies. However, BHs also power active galactic nuclei (AGNs).
In such systems, the brightness of the non-stellar continuum flux from the
AGN complicates direct measurement of the BH mass from kinematical tracers
around the central object.  However, BH masses measured in broad-line AGNs
using reverberation mapping \citep{blandfordmkee82, peterson93} have also
been found to follow a \msig\ relation
\citep{gebhardt00b,ferrarese01,onken03,nelson04}.  Reverberation mapping
relates the properties of gas in the broad-line region (BLR) to variability
in the ionizing continuum flux from the central AGN.  The time delay between
variations in the continuum and the response of a broad emission line
arising from the BLR is used to determine the characteristic radius ($r$) of
the BLR, and the width of the emission line is used to estimate the
line-of-sight velocity dispersion of the gas at that radius.
\citet{collin06} showed that measuring the line width as the second moment
of the variable part of the emission line profile ($\sigma_{\rm line}$) is
less biased than using the full width at half maximum (FWHM). The mass is
then given as $\mbh = fr\sigma_{\rm line}^{2}/G$, where $G$ is the
gravitational constant and $f$ is a scaling factor that depends on the BLR
geometry and kinematics.  Because the BLR is not spatially resolved with
current technology, mass measurements from reverberation mapping (and,
specifically, the value of $f$) must be calibrated against BH masses derived
through other methods.  One way to determine $f$ is to use the correlation
between central stellar velocity dispersion ($\sigma_{\ast}$) and $\mbh$ for
those galaxies in which reverberation mapping estimates are available.
\citet{onken04} calibrated the \msig\ relationship for 16
reverberation-mapped AGNs and calculated the average scaling factor $\langle
f\rangle = 5.5 \pm 1.8$.

Another method of calibrating the BH masses from reverberation mapping is to
apply one of the techniques commonly used for estimating BH masses in
quiescent galaxies (e.g., dynamical mass estimation from stellar kinematical
data) to a reverberation-mapped AGN. This approach is an important step in
checking the masses of BHs determined from reverberation mapping since it
does not rely on a secondary indicator of the mass (like $\sigma_{\ast}$ in
the case above).  To firmly constrain BH masses, the kinematic measurements
must reach inside the sphere of influence, the region wherein the stellar
dynamics are dominated by the BH. Satisfying this requirement is
particularly challenging in AGNs, in which, at the small angular separations
corresponding to the radius of the sphere of influence ($r_{\rm h} = G
M_{\rm BH}/\sigma_{\ast}^{2}$), the stellar light is easily overwhelmed by
the luminosity from the AGN.

Based on the revised BH masses provided by the reverberation reanalysis of
\citet{peterson04} and corresponding bulge velocity dispersions measurements
\citep{ferrarese01,onken04,nelson04}, there are two reverberation-mapped
AGNs for which the BH sphere of influence should be spatially resolvable
with existing facilities: NGC~3227 (as mentioned above) and NGC~4151, both
local broad-line Seyfert galaxies.  The BH masses in these two AGNs were
recently estimated through observations of their nuclear gas kinematics and
by modeling the gas velocity field as a flat, circular disk \citep{hicks07}.
In addition, \citet{davies06} measured the BH mass in NGC~3227 from stellar
dynamical modeling. We discuss these results further in \S~\ref{discuss}. A
handful of narrow-line AGNs also have dynamical estimates of their BH masses,
but because narrow-line objects are not capable of being reverberation mapped,
these objects cannot act as calibrators between the two mass scales.

Here we present the results of a program to measure the BH mass in NGC~4151
via stellar kinematical modeling. In \S~\ref{observations}, we describe the
imaging and spectroscopic observations as well as the data analysis process.
In \S~\ref{modeling}, we provide a brief overview of the standard 3-integral
orbital superposition method of stellar dynamical modeling that we use to
fit the data. In \S~\ref{results}, we present the results of the modeling.
In \S~\ref{discuss}, we discuss the implications of our results for the
calibration of BH masses from reverberation mapping and future steps.

\section{Observations and data analysis}
\label{observations}

\subsection{NGC~4151}

NGC~4151 is a disk galaxy with faint spiral arms, a weak large-scale bar,
and a prominent bulge. It is one of the original sample of galaxies
identified by \citet{seyfert43} as having unusual nuclear emission line
properties, and has been studied extensively over a wide range of
wavelengths \cite[see][]{ulrich00}. The heliocentric radial velocity of the
galaxy is 998~km~s$^{-1}$ \citep{pedlar92}, implying a distance of 13.9~Mpc
if in pure Hubble expansion with $H_0=72$~km~s$^{-1}$~Mpc$^{-1}$.  However,
other estimates of the distance to NGC~4151 have ranged from 10 to 30~Mpc
based on various estimates of the Virgocentric infall correction \cite[see
the discussion by][]{mundell99}. We adopted a distance of 13.9~Mpc, but note
that our dynamical estimate of the BH mass scales linearly with distance.
For example, applying the local infall correction of \citet{mould00} would
give a distance of 17~Mpc, increasing our estimated masses by roughly 20\%.
At our assumed distance of 13.9~Mpc, 1$^{\prime\prime}$ corresponds to
67~pc.

The (distance-independent) reverberation mass estimate for NGC~4151 given by
\citet{peterson04} was 1.33$\times 10^{7}$~M$_{\odot}$ (from a combination
of H$\alpha$ and H$\beta$ measurements). However, a recent reanalysis of
archival UV monitoring data from the {\em International Ultraviolet
  Explorer} \citep{metzroth06} and a new H$\beta$ reverberation campaign
\citep{bentz06b} indicate a higher BH mass: 4.57$^{+0.57}_{-0.47}\times
10^{7}$~M$_{\odot}$.  NGC~4151 has a bulge velocity dispersion of
93$\pm$14~km~s$^{-1}$ \citep{ferrarese01}, which, when combined with the
Bentz et al.\ reverberation mass, gives a sphere of influence with $r_{\rm
  h}=23$~pc or 0$.\!\!^{\prime\prime}34$.

\subsection{Imaging Data}

We made use of a set of ground-based {\em BVR} images from the MDM 1.3~m
McGraw-Hill telescope as well as {\em Hubble Space Telescope} ({\em
  HST}) imaging data obtained with the High Resolution Channel (HRC) of the
Advanced Camera for Surveys (ACS).  Both datasets are part of an independent
project designed to study host galaxy contamination in AGN spectra
\citep{bentz06}.

\subsubsection{Imaging with {\em HST}}

\begin{figure}
\epsscale{0.9}
\plotone{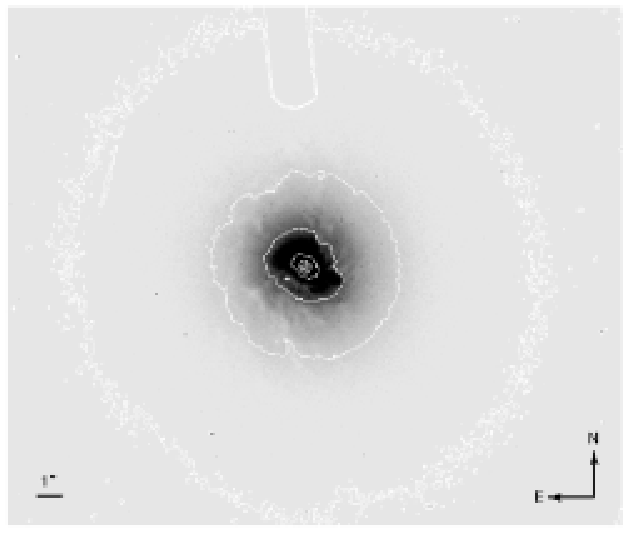}
\caption{20.5\sec$\times$24.5\sec\ portion of the {\em HST} ACS/HRC F550M
  image of NGC~4151. The logarithmic contours atop the linear grayscale show
  both the bright nuclear point source and the lower contrast features at
  larger radius. The indicated scale denotes 1~arcsec. The HRC occulting
  finger can be seen at the top of the frame.}
 \label{fig1}
\end{figure}

The ACS/HRC data were obtained with the F550M filter and have a total
exposure time of 1020~s. The full details of the data reduction are
described by \citet{bentz06}, and the final {\em HST} image is shown in
Figure~\ref{fig1}.  The image has a plate scale of
0$.\!\!^{\prime\prime}$025~pixel$^{-1}$ and a field of view of
$\sim25^{\prime\prime}\times29^{\prime\prime}$, with a point spread function
(PSF) FWHM of $\approx 0.\!\!^{\prime\prime}06$.

\subsubsection{Ground-based imaging \label{gbi}}

\begin{figure}
\epsscale{0.9} 
\plotone{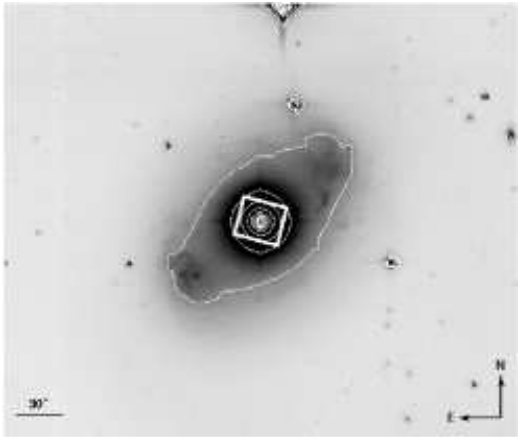}
\caption{4.5\min$\times$5.5\min\ portion of the MDM $R$-band image of
  NGC~4151.  The logarithmic contours trace the bulge and the nuclear point
  source, while the linear grayscale enhances the surface brightness
  features of the bar. The ACS/HRC field of view is overlaid as the thick
  box. The indicated scale denotes 30~arcsec.
 \label{fig2}}
\end{figure}

The MDM 1.3~m McGraw-Hill telescope was used with the ``Templeton'' CCD to
take direct images of NGC~4151 in $BVR$. The respective total exposure times
were 1470, 1200, and 1220~s. The images in each filter were bias-subtracted,
flat-fielded, registered, and then median-combined. The $R$-band image
(Fig.~\ref{fig2}) was used to determine the surface brightness profile on
large scales (\S~\ref{MGE}). The image has a plate scale of
0$.\!\!^{\prime\prime}$508~pixel$^{-1}$, a field of view of roughly
$8^{\prime}$ on a side, and a seeing FWHM of $\approx
2.\!\!^{\prime\prime}3$.  Flux calibration was achieved through aperture
photometry of a star in the image \citep{eyermann05}. 

\begin{figure}
\epsscale{0.9}
\plotone{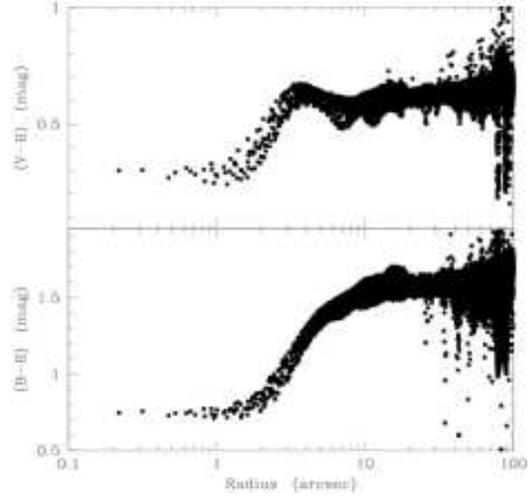}
\caption{{\em Top}: ($V-R$) color in each MDM image pixel as a function of
  NGC~4151 galactocentric radius (in arcseconds). {\em Bottom}: ($B-R$)
  pixel color as a function of radius.
\label{fig3}}
\end{figure}

\citet{bell01} employed models of stellar population synthesis and spiral
galaxy evolution, in conjunction with measured rotation curves, to derive
relationships between galaxy color and stellar mass-to-light ratio. We used
our multi-color imaging data to estimate the $R$-band stellar mass-to-light
ratio, \mlr\ (in units of solar masses per $R$-band solar luminosity
throughout), as follows. The images were aligned, convolved with Gaussians
of appropriate size to bring the PSFs to the same size as the band with the
worst seeing, and a pixel-by-pixel color map was created (Fig.~\ref{fig3}).
The ($B-R$) and ($V-R$) colors were then determined as a function of radius.
The strong decrease in ($B-R$) and ($V-R$) near the center is consistent
with AGN contamination, therefore we ignored the central 3$^{\prime\prime}$.
With ($B-R$)$\approx$1.5~mag and ($V-R$)$\approx$0.6~mag, we estimated \mlr\
to be $\approx 3$ \citep{bell01,bell03}.

\subsection{Surface Brightness Profile Modeling \label{MGE}}

We derived the surface brightness profile from the MDM $R$-band image and the
{\em HST} F550M image. As the goal of this procedure was to develop a model
for the mass distribution in the galaxy, we needed to remove the AGN
contribution to the light profile and extract the stellar surface
brightness. The process of removing the AGN light was complicated by the very
different spatial characteristics and filter bandpasses of the two images.
We used
GALFIT\footnote{\url{http://zwicky.as.arizona.edu/$\sim$cyp/work/galfit/galfit.html}},
the publicly available 2-D galaxy fitting software \citep{peng02}, to remove
the AGN contamination. In order to most accurately subtract the AGN flux, we
performed a multi-component fit to the major surface brightness features of
the galaxy and jointly determined the galaxy properties and the AGN flux to
be removed.

For the MDM $R$-band image, we subtracted a sky level equal to the median
flux beyond the visible extent of NGC~4151, and fit two S\'{e}rsic profiles
(one for the disk and one for the bulge), allowing the S\'{e}rsic index to
be a free parameter in each component. We also used a PSF for the AGN
emission that was based on one of the stellar profiles in the image. The fit
was done in several steps, initially masking out the bulge and nucleus in
order to fit the disk.  Next, we removed the mask from the bulge and used
the earlier results from the disk-only fit as the starting values for the
disk component.  Then we removed the mask from the nucleus and allowed the
nuclear magnitude to vary. We left the mask in place over several stars in
the field and over a number of bright knots near the ends of the weak bar in
NGC~4151.  Because of slight differences in the profile shapes of the
stellar PSF and the AGN light (possibly due to small image distortions
between the positions of the AGN and PSF star in the field of view), manual
adjustment of the PSF magnitude and position was required to produce the
final fit. The best-fit S\'{e}rsic index values for the disk and bulge
components were 0.73 and 1.48, respectively.

The small field of view of the {\em HST} image meant there were no
galaxy-free regions from which to measure the sky flux. Instead, the
background count rate for the HRC in the F550M filter was retrieved from the
ACS Instrument Handbook \citep{pavlovsky04} and was multiplied by the total
exposure time to estimate the background flux. The effective radii of the
two S\'{e}rsic profiles from the fit to the MDM image were converted to the
appropriate pixel scale for the {\em HST} image and held fixed. The total
magnitudes of the components were allowed to vary to account for differences
arising from the different bandpasses between the images.  We attempted to
model the HRC PSF with Tiny
Tim\footnote{\url{http://www.stsci.edu/software/tinytim/tinytim.html}},
which produces PSFs for the various {\em HST} instruments at any desired
position in the field. Tiny Tim also allows one to input a user-defined
spectrum to better match any chromatic distortions in the optical path. We
joined two archival Space Telescope Imaging Spectrograph (STIS) spectra of
NGC~4151 taken with the G430L and G750L low-resolution gratings and a
$52^{\prime\prime}\times 0.\!\!^{\prime\prime}1$ slit \citep{nelson00}, and
fed the resulting spectrum to Tiny Tim.  However, the output PSF proved to
be a poor match to the nuclear emission.  We searched for all archival
stellar images taken with the HRC in the F550M filter.  Unfortunately, this
search turned up only a handful of white dwarf observations, all with
relatively short exposure times. The best of these was a 2~s image of
BD+17$^{\circ}$4708 \citep{bohlin04}, and this image was then used as our
HRC PSF.

There were some strong residuals around the best GALFIT fit to the {\em HST}
image, but it was unclear whether they were primarily due to dust
obscuration or to [\ion{O}{3}]~$\lambda$5007 emission leaking into the F550M
bandpass.  To assess the latter possibility, we retrieved an [\ion{O}{3}]
image of NGC~4151 taken with the linear ramp filter on the Wide Field
Planetary Camera 2 (WFPC2) instrument of {\em HST} \citep{hutchings99}.
After registering the images and resampling the HRC image to the slightly
larger pixel scale of WFPC2, we were unable to find a scaling
factor for the [\ion{O}{3}] image that produced a satisfactory subtraction
of the remaining residuals. While some of the positive-flux regions in the
residual image may have been due to [\ion{O}{3}] emission, it is likely that
dust obscuration also played an important role in shaping the F550M image
morphology. Ultimately, these regions were simply included as-is in the
GALFIT fitting procedure.

\begin{figure}
\epsscale{0.9}
\plotone{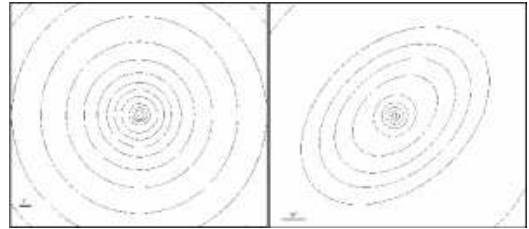}
\caption{Logarithmic contours of the AGN-free MGE model of the surface
  brightness. {\em Left}: On the size scale of the {\em HST} image (as in 
  Figure~\ref{fig1}). {\em Right}: On the size scale of the MDM image (as in 
  Figure~\ref{fig2}).
\label{mgecont}}
\end{figure}

\begin{figure}
\plotone{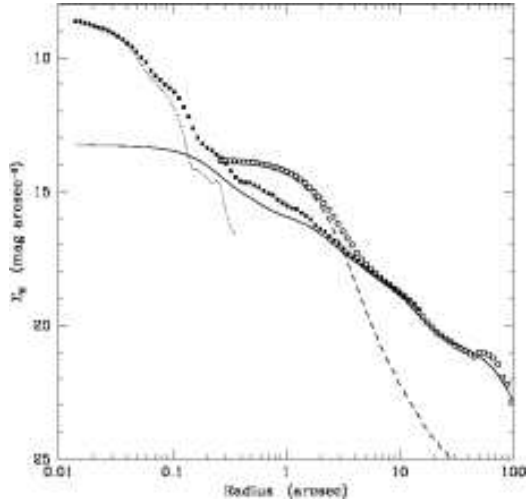}
\caption{$R$-band surface brightness along the major axis. The AGN-free profile
  determined by the MGE routine and used in the dynamical modeling is shown
  as the solid line. The raw image profile (including the AGN) and the PSF
  are shown for the {\em HST} data (filled points and dotted line) and for
  the MDM data (open points and dashed line).
  \label{figprof}}
\end{figure}

With the AGN emission subtracted, we employed the Multi-Gaussian Expansion
(MGE) routine described by \citet{cappellari02} to deproject the surface
brightness distribution into a 3-D luminosity model. The MGE routine fits a
series of $\approx 10$ concentric elliptical Gaussians to the 2-D galaxy
image. (The program has been made publicly
available\footnote{\url{http://www.strw.leidenuniv.nl/$\sim$mcappell/idl/}}
for use with IDL\footnote{\url{http://www.rsinc.com/idl/}}.)  The fit was
performed simultaneously to the MDM and {\em HST} images, and this allowed
us to rescale the {\em HST} flux level to match that of the MDM image where
the two overlapped in radius. The scaling was consistent with that expected
from the differences in plate scale and filter bandpass. We then tied the
overall flux calibration to that determined by the stellar photometry in the
$R$-band.  This approach had the advantage that the dynamical \mlr\ values
could then be compared to the \mlr\ estimates from the multi-color MDM
photometry.  Adopting an $R$-band extinction\footnote{Retrieved from NED:
  \url{http://nedwww.ipac.caltech.edu/}} in the direction of NGC~4151 of
$A_{R}$=0.074~mag \citep{schlegel98}, we converted the MGE output of flux in
each Gaussian to solar luminosities (Table~\ref{tabmge}). The 2-D contours
of the MGE model are shown in Figure~\ref{mgecont} (with the same size scale
as Figures~\ref{fig1}\ \&\ \ref{fig2}), and the major-axis $R$-band surface
brightness profiles of both the raw data (determined with the IRAF task
ELLIPSE) and the AGN-free MGE output are shown in Figure~\ref{figprof}.

\subsection{Spectroscopic Data}

We acquired spectroscopic data at the MMT Observatory, and we revisited the
Kitt Peak National Observatory (KPNO) spectra of NGC~4151 that had been used
by \citet{ferrarese01} to determine $\sigma_{\ast}$. The stellar dynamics
were determined using the \ion{Ca}{2} triplet (hereafter CaT) stellar
absorption lines at $\lambda\lambda$8498, 8542, and 8662 \AA.  We extracted
the stellar kinematics along the two (different) position angles (PAs) of
the MMT and KPNO data.

\subsubsection{MMT Spectra}

We observed NGC~4151 and velocity standard stars with the Blue Channel
Spectrograph at the 6.5~m MMT on UT 29 May 2004. We used the 1200
lines~mm$^{-1}$ grating with the LP-530 order-blocking filter, a
1$^{\prime\prime}$ slit width, and the ``ccd35'' detector. The spectra were
centered at 8600~\AA\ for measurement of the CaT with a dispersion of 0.50
\AA~pixel$^{-1}$ and a resolution of 1.4 \AA. The plate scale was
0$.\!\!^{\prime\prime}$3~pixel$^{-1}$, and the PA was 69$^{\circ}$. While
the photometric conditions were quite good, strong and gusty winds caused
both poor seeing and problematic guiding (yielding a final spatial profile
with FWHM$\sim 3^{\prime\prime}$). Our total exposure time was 11700~s.

The spectra were rectified and wavelength-calibrated using the bright sky
lines that permeate the observed spectral region (the same wavelength
solutions were applied to the spectra of a K-giant star, HR~4521, as the
stellar exposures were too short for sky lines to appear and the lines
present in the HeNeAr calibration lamp spectra were much more sparse).  Flat
field spectra taken among the observations of NGC~4151 were employed to
remove the strong fringing (20\% peak-to-peak) present in the spectra.
Although the corrections left residual fringes of nearly 10\%, the spectral
regions around the bluest two lines of the CaT were free from contamination,
and so we limited our subsequent analysis to those lines.

\subsubsection{KPNO Spectra}

The work of \citet{ferrarese01} used the Ritchey-Chr\'{e}tien Spectrograph
on the Mayall 4~m telescope at KPNO on UT 9 April 2001 to measure the bulge
velocity dispersion of NGC~4151. Details of the observations were fully
described in \citet{ferrarese01}, but we restate the key features here. The
observations were taken with a 2$^{\prime\prime}$ slit at PA=135$^{\circ}$,
the detector plate scale was 0$.\!\!^{\prime\prime}$69~pixel$^{-1}$, and the
spectral resolution was $\approx1.7$~\AA. The total exposure time was
3600~s.

The observations were rectified and wavelength-calibrated with the sky lines
present in the spectra, and then combined. Spectra of the star HR~4521 were
obtained for the absorption line profile measurements and were reduced in
the same way as the AGN data.

\subsection{Spectroscopic Analysis}

The MMT spectra were extracted in 1-pixel (0$.\!\!^{\prime\prime}$3)
apertures out to $\pm 12^{\prime\prime}$ ($\pm$40 pixels), tracing along the
peak of the spatial profile. While the seeing disk (modeled as a Gaussian
with $\sigma=1.\!\!^{\prime\prime}27$) was much larger than our individual
extraction windows, the dynamical modeling routine took the seeing into
account when fitting to the observed kinematics. To isolate the CaT lines
for profile fitting, the extracted spectra were normalized with a high-order
spline fit to the continuum, any remaining fringing, and several AGN
emission features.

Because of the lower signal-to-noise ratio, $S/N$, in the KPNO data,
extractions were made with a variable window size, ranging from 1 pixel
(0$.\!\!^{\prime\prime}$69) at the peak of the light to a 7-pixel width in
the furthest bins (extending to distances of $\approx$20 pixels, or
$14^{\prime\prime}$). The seeing for these observations was modeled as a
Gaussian with $\sigma=0.\!\!^{\prime\prime}76$.

The spectroscopic data were analyzed with the Penalized Pixel-Fitting (pPXF)
method of \citet{cappellari04}\footnote{The IDL routine is publicly
  available from \url{http://www.strw.leidenuniv.nl/$\sim$mcappell/idl/}.}.
The pPXF program finds the best fit to the line-of-sight velocity
distribution (LOSVD) based on a maximum penalized likelihood approach to the
broadening of a template spectrum. The method parametrizes the LOSVD in the
standard way as a Gauss-Hermite series \citep{vandermarel93, gerhard93},
which we limited in our analysis to the first four terms ($V$, $\sigma$,
$h_3$, $h_4$), where $V$ is the radial velocity, $\sigma$ is the velocity
dispersion, and $h_3$ and $h_4$ are the coefficients of the next two
elements of the Gauss-Hermite expansion.

\begin{figure}
\plotone{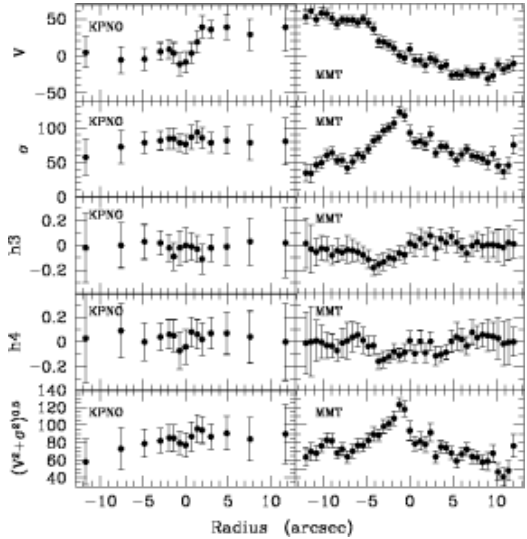}
\caption{Kinematic data from KPNO ({\em left}) and MMT ({\em right})
  parameterized as Gauss-Hermite coefficients $V$, $\sigma$, $h_3$, and
  $h_4$, and the rms velocity ($\sqrt{V^2+\sigma^2}$) as functions of
  spectroscopic slit position.
  \label{kinemfig}}
\end{figure}

The pPXF routine uses a stellar spectrum as its template and simultaneously
fits to all desired Gauss-Hermite terms over a user-defined wavelength
region. The ability to select multiple fitting regions allowed us to exclude
regions of noise or AGN emission that might lie between the stellar
absorption lines we are trying to measure. Each of the extracted spectra
from the MMT and KPNO data were run through the pPXF algorithm with a
corresponding stellar template. HR~4521 was the only template common to both
datasets. Thus, to minimize systematic differences between the MMT and KPNO
results, and because the CaT is generally insensitive to problems from
mismatched stellar templates \citep{barth02}, we limited our analysis to the
single template star. The resulting kinematic parameters are shown in
Figure~\ref{kinemfig} and listed in Table~\ref{tabkin}.

In addition to the narrow-aperture extractions described above, the MMT
spectrum was extracted with a width of $\approx 4^{\prime\prime}$ to provide
an independent measurement of the bulge velocity dispersion. Applying the
pPXF procedure gave $\sigma_{\ast}=109\pm11$~km~s$^{-1}$, reasonably
consistent with the results of both Ferrarese et al.\ (2001;
$93\pm14$~km$^{-1}$) and Nelson et al.\ (2004; $97\pm3$~km~s$^{-1}$).

\subsection{Target-of-Opportunity Observations with STIS}

One way to minimize the problem of nuclear glare when attempting to measure
stellar features in an AGN is to take advantage of the variable nature of
the nuclear emission and observe the target when the AGN is dim. We obtained
Target-of-Opportunity (ToO) observations of NGC~4151 with STIS on {\em HST}.
Monitoring of both NGC~4151 and NGC~3227 was carried out with ground-based
spectroscopy at the Crimean Astrophysical Observatory (CrAO) and with
space-based X-ray flux measurements from the {\em Rossi X-ray Timing
  Explorer} ({\em RXTE}).

The threshold for triggering the ToO observation had been derived from
ground-based $R$-band observations and bulge-disk-nucleus decompositions of
NGC~3227 and NGC~4151 \citep{virani00}.  These data allowed us to estimate
the stellar flux we could expect to fall within the STIS slit and also to
independently examine the effects on the $S/N$ of varying the nuclear
brightness. We assumed a stellar contribution near the center of each galaxy
to be roughly like an elliptical galaxy (flat in $F_{\lambda}$), with a
spatial distribution given by the bulge fit of \citet{virani00}. Thus, we
were able to parametrize the exposure time as a function of the nuclear
non-stellar flux. We calculated the thresholds of nuclear flux for each
target that allowed us to reach our desired stellar $S/N$ of 50 \AA$^{-1}$
in a reasonable amount of exposure time. For NGC~4151, this threshold was
equal to $4\times 10^{-14}$~erg~s$^{-1}$~cm$^{-2}$~\AA$^{-1}$, which the AGN
had been near in the recent past.

\begin{figure}
\epsscale{0.9}
\plotone{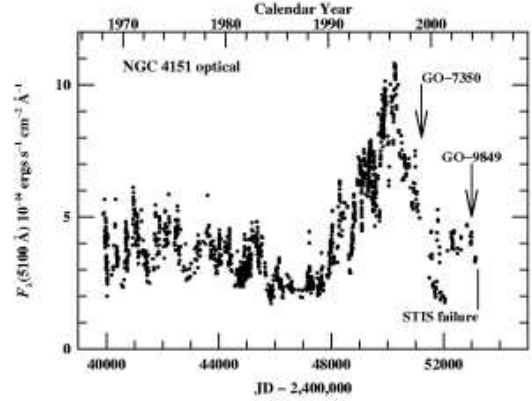}
\caption{Historical light curve of optical continuum flux in NGC~4151. The dates of an
  earlier STIS program that used the occulting bar (GO-7350), our ToO
  observations (GO-9849), and when STIS stopped functioning are indicated.
  \label{fig4}}
\end{figure}

On 2003 November 19, the CrAO observations of NGC~4151 indicated that the
continuum flux level had fallen below the triggering threshold.
Figure~\ref{fig4} shows a historical light curve for NGC~4151 and indicates
with arrows the date of a previous STIS observation (a shorter exposure that
employed the STIS occulting bar to block the nuclear light, but which also
blocked the stellar signature near the BH), the date when our ToO
observation began, and when the STIS instrument stopped functioning.

The STIS observations were executed on UT 2003 December 15 and 19. The
instrumental setup consisted of the G750M grating and the
52$^{\prime\prime}\times 0.\!\!^{\prime\prime}1$ slit, oriented with a PA of
69$^{\circ}$.  The total exposure time was 29730~s. In addition to the
observations of NGC~4151, spectra of HR~4521 were obtained on UT 2003
December 26 with the same instrumental setup to act as our comparison for
the CaT line profiles.

While the continuum flux stayed close to the threshold value between the ToO
triggering on 2003 November 20 and the actual time of observation, the
extracted STIS spectrum does not show the presence of the CaT lines, and so
we were unable to use the STIS data to constrain the stellar kinematics near
the BH in NGC~4151.

The lack of absorption lines can be used to place upper limits on the number
of giant stars in the central region of NGC~4151. The spectrum of the
comparison star HR~4521 was used to estimate the maximum number of such
stars that could exist in the center of NGC~4151 without producing
measurable lines. HR~4521 has a {\em Hipparcos} parallax measurement of
15.80$\pm$0.59~mas, implying a distance of 63.3$\pm$2.4~pc. We then scaled
the luminosity of HR~4521 to the flux appropriate for our assumed distance
of NGC~4151 (13.9~Mpc), and found the multiplicative factors required to
produce visible spectral features when subtracting the scaled stellar
spectra from NGC~4151 spectra of different aperture sizes. For the inner
0$.\!\!^{\prime\prime}$1 (6.7~pc), we derived an upper limit of $7.5\times
10^5$ giant stars. A wide-aperture extraction ($\sim 3^{\prime\prime}\times
0.\!\!^{\prime\prime}1$, or $\sim$200~pc $\times$ 6.7~pc) only increased the
upper limit to 1.25$\times 10^6$ stars. For comparison, the mass enclosed
within a radius of 7~pc at the center of the Milky Way is $\sim2\times
10^{7}$ M$_{\odot}$, only about 10\% of which is accounted for by the BH
\cite[see][]{melia01}. Approximately $\sim$1\% of the stellar mass within
that region of the Milky Way is composed of bright giants \citep{genzel96},
giving roughly 2$\times 10^5$ M$_{\odot}$ and the equivalent number of stars
(assuming an old bulge stellar population). Whether we take the NGC~4151
bulge mass estimate of Wandel (2002; a factor of $\sim$4 smaller than the
Milky Way bulge mass given by Bissantz et al.\ 1997), or we estimate the
bulge mass from the $\mbh$-$M_{\rm bulge}$ relation of H\"{a}ring \& Rix
(2004; a factor of 3 larger than the Milky Way bulge), the lack of CaT
absorption is consistent with NGC~4151 having a similar bright giant
fraction as the Milky Way.

As noted by \citet{bentz06}, ground-based imaging data often are
insufficient to make an accurate galaxy decomposition. Thus, our reliance on
the ground-based imaging analysis of \citet{virani00} to estimate the
appropriate exposure time for the STIS observation may have contributed to
our failure to detect the stellar absorption features. However, it is also
possible that the CaT lines were filled in by emission arising from either
young stars near the center of NGC~4151 or from the AGN itself.

\section{Dynamical Modeling \label{modeling}}

We carried out dynamical modeling of the kinematical and imaging data using
the now standard method of orbit superposition first described by
\citet{schwarzschild79}. This allowed us to set dynamical constraints on the
possible values of the BH mass ($\mbh$) and the stellar mass-to-light ratio
(\mlr) that are consistent with the surface brightness and spectral
data.  The modeling algorithm we used is the one described by Valluri et
al.\ (2004; hereafter VME04) and we refer the reader to that paper for a
detailed description of the method.

We constructed a 3-D mass model of the galaxy by deprojecting the surface
brightness distribution produced by the MGE routine (\S~\ref{MGE}), using
the method outlined by \citet{cappellari02}, and assuming an error of
$10^{-4}$ in each mass cell. We also assumed an inclination angle of the
bulge (described below) and, to normalize the mass density, a value for
\mlr, which was constant with radius.  An unknown point mass ($\mbh$)
representing the central supermassive BH was added to the mass distribution
and a large set of orbits were integrated in the resultant gravitational
potential for every pair of parameters ($\mbh$, \mlr), building an orbit
library for each ($\mbh$, \mlr) combination.  We constructed models for a
wide range of values of $\mbh$ and \mlr\ (12--17 different values of $\mbh$
and 8--16 values of \mlr).

Two different inclination angles were used to model the bulge.  The
large-scale galaxy disk is somewhat inclined to the line of sight \cite[the
rotation axis of the disk having an inclination of $i=20-25^{\circ}$
relative to the line of sight;][]{simkin75}.  However, the bulge appears
very close to circular in projection (axial ratio $q\sim 0.99$), suggesting
that it is nearly spherical. We therefore constructed two sets of models. We
first assumed that the bulge is viewed edge-on ($i=90^{\circ}$) and has its
kinematic symmetry axis in the plane of the sky.  The assumption of edge-on
axisymmetry should be reasonable for orbits near the BH.  We also
constructed a second set of models for which we assumed that the rotation
axis of the axisymmetric bulge has an inclination of $i=23^{\circ}$.
Because of these motivations for the two particular values of the
inclination, we treat the cases independently, rather than marginalizing
over the inclination.

The presence of a weak, large-scale bar in NGC~4151 is noticeable beyond
50$^{\prime\prime}$ and is expected to result in non-axisymmetric motions in
the stellar orbits, which cannot be modeled by our axisymmetric code.
Although the bar is expected to be kinematically weak \citep{ms99}, we
restricted our models to fit the light distribution between
0$.\!\!^{\prime\prime}$00015 (0.01~pc) and 50$^{\prime\prime}$ (3.35~kpc).
Again, within this radius, the projected isophotes were almost circular.

The total number of luminosity constraints from the surface brightness
profile was 266 (the inner 196 from the {\em HST} photometry, and the
remainder from the MDM photometry).  Kinematical constraints ($V, \sigma,
h_3, h_4$) were available for 58 apertures within $\sim 15$\sec\, from the
KPNO and MMT observations.  Since the kinematical quantities (especially $V$
and $\sigma$) obtained with both spectrographs were not symmetrical with
respect to the center of the galaxy, the data at each slit position $+x$
were averaged with those from $-x$, as is standard procedure.  The
Schwarzschild orbit superposition code assumes that the mass distribution is
axisymmetrical.  Consequently when the mass contribution of an orbit is
stored in the library, it is stored on a grid in the meridional plane of the
galaxy in cylindrical polar coordinates. However since we fit the
kinematical data in the plane of the sky, the kinematical information from
the orbit library was stored in the plane of the sky. It is necessary to
symmetrize the kinematical data because if one does not do so, the model
will try to simultaneously fit non-axisymmetric kinematical features and an
axisymmetric mass distribution.  Since these conditions are physically
inconsistent with each other, the solutions obtained have poor $\chi^2$ and
are biased.  The total number of kinematical constraints modeled were 4
Gauss-Hermite parameters $\times$ 29 apertures$ = 116$.

Each model was constructed with an orbit library consisting of 8100
different orbits.  The contributions of each orbit to the observed LOSVDs
and the deprojected luminosity distribution were determined. A non-negative
least squares programming routine \citep{lawson95} was used to find the
weighted superposition of the orbits that best reproduced both the assumed
model stellar density distribution and the observed kinematical data.

Smoothing, or ``regularization'', of the orbital solutions is typically
performed in this type of analysis, and a variety of methods have been
employed to that end. One can locally smooth the phase space
\citep{cretton99}, impose a ``maximum entropy'' constraint
\citep{gebhardt03}, or include a penalty function with an adjustable
smoothing parameter (VME04). In the case of the penalty function, poorly
selected values of the smoothing parameter can generate misleading results
(VME04). For example, over-smoothing effectively restricts the range of
allowed orbits, and can give erroneous best-fit values of $\mbh$. Here we
present models without regularization. From VME04, we know that models
without regularization give the full range of parameters allowed by the
data.  Regularization mainly helps to reduce the sensitivity of the
solutions to noise in the data by requiring a smoother sampling of the orbit
libraries used in the solution. However, when the unregularized solutions
give a poor fit to the data, regularization does not improve the constraints
obtained on the parameters.

Without regularization, the number of orbits used in the fits is generally
equal to the number of constraints. In the edge-on model, orbit libraries
employing half the number of orbits produced no difference in the results.
In contrast (as described below), small changes in the orbit library size
for the inclined model indicated some sensitivity to the number of orbits
used, although not in a systematic manner. Since the number of orbits
available influences the uncertainties in the solution (VME04), we analyzed
both the inclined and edge-on cases with 8100 orbits, providing the most
uniform comparison of the models.

\section{ Results of 3-Integral Modeling \label{results}}

\begin{figure}
\epsscale{0.9}
\plotone{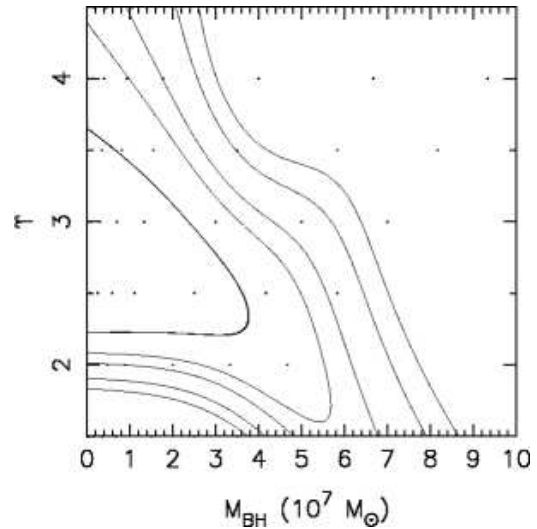}
\caption{Contours of constant $\chi^2$ derived by fitting the kinematical
  data with 3-integral axisymmetric models for a bulge that is assumed to be
  viewed edge-on. The dots correspond to the values of parameters
  ($\mbh$,\mlr) for which solutions were obtained. The three inner
  most contours (moving left to right at \mlr $\sim$ 3) correspond to the
  1$\sigma$, 2$\sigma$, and 3$\sigma$ confidence intervals respectively 
  (68.5\%, 90\% and 95\%). The contour plot is obtained using a smoothing 
  kernel.}
\label{chi2d90}
\end{figure}

\begin{figure}
\epsscale{0.9}
\plotone{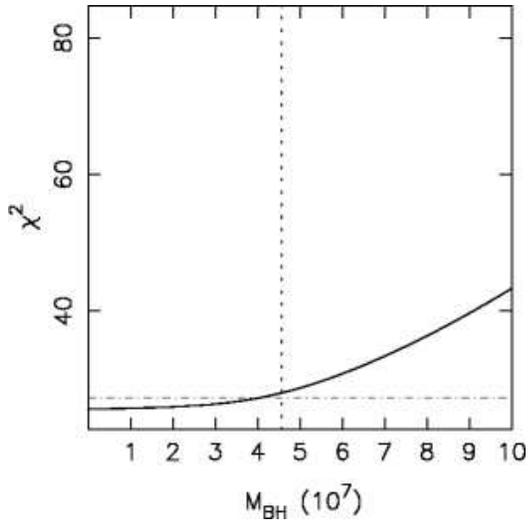}
\caption{$\chi^2$ values as a function of $\mbh$ obtained by taking a cut
  through Figure~\ref{chi2d90} at \mlr\ = 2.3. The horizontal dot-dash
  line is at $\Delta \chi^2 = 2.3$ (68.5\% confidence interval for 2 DoF)
  above the minimum value of $\chi^2$. The vertical dashed line indicates
  the best-fit value obtained from reverberation mapping.}
\label{chi1d90}
\end{figure}

We present the results of 3-integral modeling with orbit libraries
consisting of 8100 orbits each, which were fit to all of the available
kinematic and photometric data.  Figure~\ref{chi2d90} shows 2-dimensional
contour plots of the total $\chi^2$ as a function of $\mbh$ (abscissa) and
\mlr\ for models which assumed that the bulge is viewed edge-on
($i=90^{\circ}$). The models covered 12 values of $\mbh$ and 8 values of
\mlr\ (indicated by the grid of points); the inner 5 contours are plotted at
$\Delta(\chi^2) = 2.30, 4.61, 6.17, 9.21, 11.8$ which correspond to 68.3\%,
90\%, 95.4\%, 99\%, 99.73\% confidence intervals, assuming two degrees of
freedom (DoF).  The innermost contour gives a 1$\sigma$ upper limit of
$4.2\times 10^7$~M$_{\odot}$ (for 2 DoF) but no best-fit value of the BH
mass, since the values of $\chi^2$ are all comparable at smaller masses.
The value of the mass-to-light ratio corresponding to the mass upper limit
is \mlr\ = 2.3, roughly consistent with the estimate from the ground-based
photometry in \S~\ref{gbi}.  Figure~\ref{chi1d90} shows a 1-dimensional cut
through the 2-dimensional contour plot at the value of \mlr\ = 2.3. The
horizontal dot-dash line is at $\Delta \chi^2 = 2.30$, which corresponds to
the $1\sigma$ confidence interval for two DoF. The vertical dashed line in
Figure~\ref{chi1d90} is at the location of the best-fit BH mass obtained
from reverberation mapping studies: $4.57^{+0.57}_{-0.47}\times
10^{7}$~M$_{\odot}$ \citep{bentz06b}.

\begin{figure}
\epsscale{0.9}
\plotone{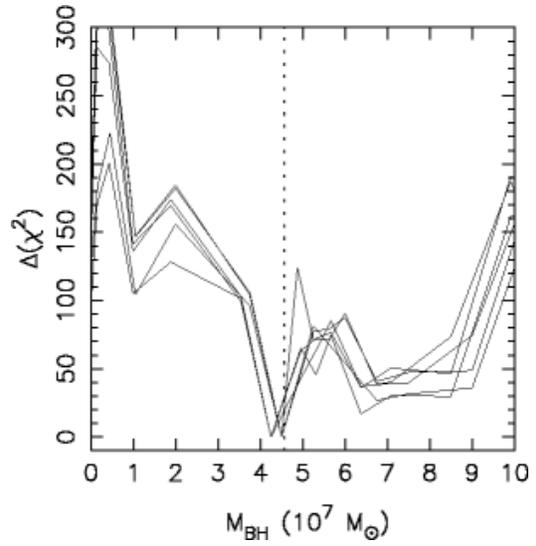}
\caption{$\chi^2$ values as a function of $\mbh$ obtained by taking cuts at
  \mlr\ =3.75 through several dynamical models with inclined bulges that
  are created with slight variations in the number of kinematic orbits
  (7600--8100). The vertical dashed line indicates the best-fit value
  obtained from reverberation mapping.}
\label{1dmulti}
\end{figure}

Figure~\ref{1dmulti} shows a similar 1-dimensional cut (at the best-fit
\mlr\ = 3.5) through the corresponding 2-dimensional $\chi^2$ space for the
models that assumed the bulge is viewed at the same inclination as the
large-scale galaxy disk (as inferred from the shape of the disk,
$i=23^{\circ}$).  As an additional test of numerical effects, we ran several
versions of the inclined model with orbit libraries of slightly different
sizes (ranging between 7600 and 8100 orbits), for which we also show the
1-dimensional cuts in Figure~\ref{1dmulti}.  There is clearly a minimum in
$\chi^2$ between $4\times 10^7$ and $5\times 10^7$~M$_{\odot}$, but the
precise location is uncertain. Furthermore, the non-smooth nature of the
$\chi^2$ topology suggests that it is unwise to assign much meaning to the
formal $1\sigma$ $\chi^2$ intervals since the location of the minimum varies
quite strongly with orbit library. The most conservative statement that can
be made is that the models give a best-fit BH mass between $4-5\times
10^7$~\msun.

However, there are two reasons why we are reluctant to accept this as more
than preliminary evidence of a ``BH detection'' based on stellar dynamical
modeling. First, the best-fit $\chi^2$ value for the inclined model is a
factor of 3 larger than the lowest value of $\chi^2$ for the edge-on model
(with the upper limit).  Formally, therefore, the solution with an edge-on
bulge and an upper-limit $\mbh$ of $4.2\times 10^7$~M$_{\odot}$ is
preferred.  Second, there is considerably more noise in the topology of the
2-D $\chi^2$ surface for the inclined models than for the edge-on models,
which is indicative of the fact that the inclined models are significantly
more influenced by the noise in the data than are the edge-on models.

\begin{figure}
\epsscale{1.}
\plotone{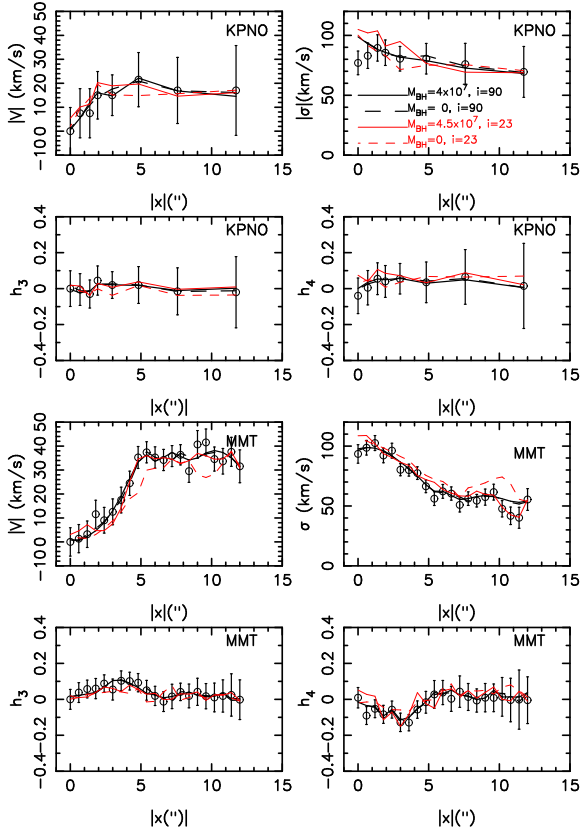}
\caption{Fit to kinematical data parametrized as Gauss-Hermite coefficients
  $V, \sigma, h_3, h_4$ as a function of position along the slit $|x|$ for
  ground-based data from KPNO (top four panels) and MMT (bottom four
  panels).  Solutions used libraries of 8100 orbits for 4 different values
  of $\mbh$ as indicated by the labels. The black curves are fits obtained
  with two edge-on models: $\mbh = 4\times 10^7$~\msun\ (solid line) and
  $\mbh = 0$ (dashed line). The red curves are for models with inclination
  angle $i=23^\circ$: $\mbh = 4.5\times 10^7$~\msun\ (solid line) and $\mbh
  = 0$ (dashed line).}
\label{N4151-kinem}
\end{figure}

Figure~\ref{N4151-kinem} shows fits for 4 different models to the
kinematical data from the two ground-based datasets (shown by open circles
with $1\sigma$ error bars). The black lines are the fits obtained for an
edge-on bulge (the solid black line is for the model at the upper limit, and
the dashed black line is for an edge-on bulge with no black hole).  The red
lines are for the models with inclination angle $i = 23^\circ$. The solid
red line is for the model that gives the best fit to the kinematical data,
while the dashed red line is for the inclined bulge with no BH.

It is clear that in the case of the edge-on models (black lines), the upper
limit and the ``no BH'' models provide essentially identical fits to the
data. For an inclined bulge, the model with no BH systematically
under-predicts the velocity ($V$), which is why the larger BH mass is needed
to fit the kinematical data.

If we make the plausible assumption that the large-scale inclination of the
disk and its symmetry axis are likely to be shared by the bulge, one can
reasonably conclude that the best-fit BH mass is
$4-5\times10^7$~M$_{\odot}$, in extremely good agreement with the estimate
obtained from reverberation mapping. Models in which the bulge grows via
secular processes or by dissipative minor mergers of disk galaxies predict
such alignment of symmetry axes \cite[e.g.,][]{robertson06}.

We propose this conclusion with some caution, however, because the
ground-based kinematical data are of relatively low quality and do not
resolve the sphere of influence of the BH.  Previous studies
\citep{verolme02} have shown that it is essential to have large-scale
two-dimensional kinematical data from integral field spectroscopy in order
to obtain reliable constraints on the inclination angle of an ellipsoidal
galaxy model from the kinematical data alone. 

The existing data and the dynamical models presented above show that there
is tentative, but good agreement between the mass of the BH inferred from
stellar dynamical models and the mass obtained from reverberation mapping.
 
\section{Discussion \label{discuss}}

We have presented one of the first direct stellar dynamical BH mass
estimates for a broad-line AGN. We found that if the bulge of NGC~4151 was
assumed to be viewed edge-on, the data only gave an upper limit on the
dynamical BH mass of $4.2\times 10^7$~\msun. Alternatively, when the bulge
was assumed to have the same inclination to the line of sight as the
rotation axis of the large-scale disk, we obtained a best-fit value for the
mass of the central BH of $\mbh = 4 - 5\times 10^7$~\msun.  The only other
stellar dynamical measurement of a broad-line AGN is from a recent study of
NGC~3227 by \citet{davies06}, who estimated a 1$\sigma$ range for the BH
mass of $7-20\times 10^{6}$~\msun.  The value of $\mbh$ derived from stellar
kinematical data and 3-integral modeling is remarkably consistent with the
value derived from reverberation mapping in both cases.  

In addition, \citet{hicks07} have analyzed the dynamics of the molecular
hydrogen gas in the central regions of both NGC~3227 and NGC~4151, and
derive BH mass estimates for these galaxies that agree well with the
reverberation mapping estimates. All of this supports the empirical
calibration of reverberation-based masses from \citet{onken04}, and
indicates that the local AGNs used to establish that calibration likely
already lie very close to the \msig\ relation defined by quiescent galaxies.
However, there is a fair amount of scatter in the AGN \msig\ relation, and
the mass predicted by the relation for NGC~4151 is $\approx 5\times
10^{6}$~M$_{\odot}$ \cite[using $\sigma_{\ast}=98$~km~s$^{-1}$ and the fit
to the relation by][]{ferrarese05}, roughly an order of magnitude lower than
suggested by either the reverberation mapping studies or the stellar
dynamical data presented here. While one could attribute the discrepancy of
the reverberation-based mass in NGC~4151 to object-to-object variations in
the reverberation scaling factor $f$, a higher value of $\mbh$ is supported
by the stellar dynamical estimate, the gas dynamical estimate, and the
relationship between BH mass, bolometric luminosity, and X-ray variability
timescale \citep{mchardy06}. We note that our stellar dynamical models for
NGC~4151 are based on kinematical data that do not fully resolve the sphere
of influence of the BH and that our modeling is simplistic in that it
assumes an axisymmetric bulge and takes no account of the non-axisymmetry
arising due to the presence of the galaxy's bar. Thus we consider this
measurement of the BH mass to be preliminary.

We believe that NGC~4151 is an excellent target for future stellar dynamical
measurements with higher spatial resolution.  One improvement over our
present approach would be to expand from two slit positions to many, or to
use an integral field unit (IFU). In the seeing-limited regime, observations
with the optical IFU, SAURON, have recently been published by
\citet{dumas07} and show a similar decrease in the stellar velocity
dispersion of NGC~4151 at small radii.  While the current shutdown of STIS
precludes deeper studies of the CaT with {\em HST}, observations of the CO
bandhead stellar absorption features at 2.29 $\mu$m \cite[as was used in the
study of NGC~3227 by][]{davies06} opens the possibility of ground-based
observations with adaptive optics (AO) on large-aperture telescopes. It is
worrying that observations of NGC~4151 taken during periods of both high AGN
flux \citep{ivanov00} and low AGN flux \citep{hicks07} fail to clearly
reveal CO bandhead absorption from the central regions of the galaxy, but
deeper observations may improve the situation.  Incorporating these types of
data into future modeling will be a significant step forward.

The use of high-spatial-resolution AO observations with infrared IFUs in
conjunction with non-axisymmetric modeling will allow us to constrain the
complex velocity fields arising from the bar; this is likely to place even
stronger constraints on the mass of the central BH. Ultimately, the
insensitivity of stellar orbits to the outflows and bulk motions that affect
gas dynamics means that stellar dynamical BH masses will likely act as the
linchpin in calibrating the BH masses produced by reverberation mapping.

\acknowledgements

This work made extensive use of software that had been publicly released. In
particular, we thank Michele Cappellari and Chien Peng for the programs they
have provided to the astronomical community. We thank the referee for their
many helpful comments. We are grateful for support of various aspects of
this work by NASA through grants HST-GO-09849 and HST-GO-09851 from the
Space Telescope Science Institute, which is operated by the Association of
Universities for Research in Astronomy, Incorporated, under NASA contract
NAS5-26555; by the National Science Foundation (NSF) through grants
AST-0205964 and AST-0604066; and by the Civilian Research and Development
Foundation through grant UP1-2549.  Bentz was supported by a Graduate
Fellowship of the NSF.  Merritt acknowledges support through grants NSF
AST-0420920, NSF AST-0437519, and NASA NNG04GJ48G. Valluri was supported by
supported by the Kavli Institute for Cosmological Physics through the grant
NSF PHY-0114422.  Vestergaard is grateful for support of this research by
NASA through grant HST-AR-10691-01.A from the Space Telescope Science
Institute, and by the NSF through grant AST-0307384. This research has made
use of the NASA/IPAC Extragalactic Database (NED) which is operated by the
Jet Propulsion Laboratory, California Institute of Technology, under
contract with the National Aeronautics and Space Administration.

%\clearpage

%\clearpage

%\clearpage

%\clearpage

%\clearpage

%\clearpage

%\clearpage

%\clearpage

%\clearpage

%\clearpage

%\clearpage

\clearpage

\begin{deluxetable}{ccrc}
\tablecaption{AGN-Subtracted Multi-Gaussian Expansion Results\label{tabmge}}
\tablewidth{0pt}
\tablehead{
\colhead{Gaussian} & \colhead{Surface Density} & \colhead{Gaussian Sigma} & \colhead{Axial}\\
\colhead{Number} & \colhead{(L$_{\odot,R}$~pc$^{-2}$)} & \colhead{(arcsec)} & \colhead{Ratio}
}
\startdata
1 & 8.75E+04 &   0.14 & 1.00 \\
2 & 2.43E+04 &   0.32 & 1.00 \\
3 & 8.38E+03 &   1.19 & 1.00 \\
4 & 3.19E+03 &   2.44 & 1.00 \\
5 & 1.25E+03 &   6.46 & 1.00 \\
6 & 2.75E+02 &  11.57 & 1.00 \\
7 & 1.50E+02 &  45.21 & 0.67 \\
8 & 5.06E+00 & 203.53 & 1.00 \\
9 & 0.88E+00 & 203.53 & 0.27 \\
\enddata
\end{deluxetable}

\begin{deluxetable}{cccccc}
\tablecaption{Gauss-Hermite Fits to Kinematic Data\label{tabkin}} 
\tablewidth{0pt} 
\tablehead{
\colhead {} & \colhead{Position} & \colhead{} & \colhead{} & \colhead{} & \colhead{}\\
\colhead {Dataset} & \colhead{Along Slit\tablenotemark{a}} & \colhead{Velocity} & \colhead{$\sigma$} & \colhead{$h_3$} & \colhead{$h_4$}\\
\colhead {} & \colhead{(arcsec)} & \colhead{(km s$^{-1}$)} & \colhead{(km s$^{-1}$)} & \colhead{} & \colhead{} 
} 
\startdata
MMT & 12.0$\pm$0.3 & -10$\pm$11 & 76$\pm$13 & 0.01$\pm$0.11 & 0.00$\pm$0.12\\
\nodata & 11.4$\pm$0.3 & -14$\pm$8 & 46$\pm$11 & 0.02$\pm$0.14 & 0.00$\pm$0.18\\
\nodata & 10.8$\pm$0.3 & -18$\pm$8 & 37$\pm$11 & -0.02$\pm$0.14 & -0.02$\pm$0.18\\
\nodata & 10.2$\pm$0.3 & -11$\pm$8 & 45$\pm$10 & 0.00$\pm$0.13 & 0.03$\pm$0.17\\
\nodata & 9.6$\pm$0.3 & -26$\pm$8 & 63$\pm$10 & 0.00$\pm$0.10 & 0.05$\pm$0.11\\
\enddata 
\tablenotetext{a}{Uncertainties indicate aperture size along the slit.}
\tablecomments{The kinematic data were symmetrized in the subsequent
  dynamical modeling. [The complete version of this table is in the
  electronic edition of the Journal. The printed edition contains only a
  sample.]}
\end{deluxetable}

\end{document}